\documentstyle[epsfig]{aipproc}

\newcommand{\CaW}{\mbox{$\mbox{CaWO}_4$}}
\newcommand{\CoS}{\mbox{${}^{60}\mbox{Co}$}}
\newcommand{\CoF}{\mbox{${}^{57}\mbox{Co}$}}

\begin{document}

\title{The CRESST Experiment:\\ Recent Results and Prospects}

\author{P.~Di~Stefano$^a$\footnote{distefano@mppmu.mpg.de}, M.~Bruckmayer$^a$, C.~Bucci$^c$, S.~Cooper$^b$, C.~Cozzini$^a$, F.~von~Feilitzsch$^d$, T.~Frank$^a$, D.~Hauff$^a$, T.~Jagemann$^d$, J.~Jochum$^d$, R.~Keeling$^b$, H.~Kraus$^b$, J.~Marchese$^b$, D.~Pergolesi$^a$, F.~Pr\"obst$^a$, Y.~Ramachers$^b$, J.~Schnagl$^d$, W.~Seidel$^a$, 
I.~Sergeyev$^a$, M.~Stark$^d$, L.~Stodolsky$^a$, S.~Uchaikin$^a$, H.~Wulandari$^d$}
\address{ $^a$Max-Planck-Institut f\"ur Physik, F\"ohringer Ring 6, D-80805 Munich, Germany\\$^b$University of Oxford, Department of Physics, Oxford OX1 3RH, UK\\$^c$Laboratori Nazionali del Gran Sasso, I-67010 Assergi, Italy\\$^d$Technische Universit\"at M\"unchen, Physik Department, D-85747 Munich, Germany}

\maketitle

\begin{abstract}
The CRESST experiment seeks hypothetical 
WIMP particles that could account for the bulk of dark matter in the Universe.  The detectors are cryogenic calorimeters in which WIMPs would scatter elastically on nuclei, releasing phonons.  The first phase of the experiment has successfully deployed several 262~g sapphire devices in the Gran Sasso underground laboratories.  A main source of background has been identified as microscopic mechanical fracturing of the crystals, and has been eliminated, improving the background rate by up to three orders of magnitude at low energies, leaving a rate close to one count per day per kg and per keV above 10~keV recoil energy.  This background  now appears to be dominated by radioactivity, 
and future CRESST scintillating calorimeters which simultaneously measure light and phonons will allow rejection of a great part  of it.
\end{abstract}

\section*{Introduction}
The question of dark matter was first raised  by astronomer Fritz Zwicky in a 1933 study of redshifts of galaxies in clusters~\cite{distefano:zwicky1933}.  
In the Coma system,
Zwicky found the mean mass density to be at least 400 times greater than the luminous mass density, 
which led him to postulate the presence of dark matter.  
Today, the problem remains, at many additional scales in the Universe, and there is a compelling cosmological case for the presence of non-baryonic weakly interacting massive particles (WIMPs) in our galactic halo~\cite{distefano:generaldm}.  WIMPs are all the more appealing that they arise in supersymmetric theories in the guise of the lightest supersymmetric particle (LSP), which is typically the neutralino.

The direct approach to detecting these halo WIMPs is to look for their interactions in a detector, as was first proposed in connection with cryogenic devices~\cite{distefano:leo,distefano:goodman}.  
The low energies (of the order of keVs) and rates (of the order of counts per day per kg of detector) involved present a major experimental challenge.   
Historically, the first detectors used were semi-conducting devices, followed by scintillating crystals and more recently, after much development, the cryogenic techniques.

\section*{CRESST Sapphire Calorimeters}
In its first phase, 
the Cryogenic Rare Event Search with Superconducting Thermometers (CRESST) employs
calorimeters 
composed of a 262~g sapphire absorber in which an incident particle creates a nuclear or an electron recoil which in turn produces non-thermal phonons.  
These phonons make their way to a thin tungsten film deposited on the crystal and maintained in its superconducting transition in the vicinity of 15~mK~\cite{distefano:colling1995}.  
The arrival of the phonons heats the film slightly, inducing a relatively large  change in its resistance~\cite{distefano:proebst1995}.  
This change is measured with a SQUID, by passing a constant current through a parallel circuit, one branch of which is the film, 
and the other a resistor and the inductance coil of the SQUID.

Typical tungsten transition curves are not always smooth, therefore reconstructing  the energy scale is not necessarily straightforward.  
CRESST calorimeters are designed to facilitate this task : they have a resistive gold wire bonded to the film, 
through which it is possible to apply a  heat pulse to the thermometer, 
mimicking  
the shape of
pulses coming from the sapphire absorber.  These are employed to inject pulses of known energy into the film, and thereby reconstruct the energy scale and  monitor trigger efficiency at low energies.  Supplied with constant current, such gold heaters can also serve to stabilize the operating point on the transition curve.
Performances of this type of detector  include
a resolution FWHM of 250~eV at 1.5~keV~\cite{distefano:sisti1999}, which can be improved to 130~eV using active thermal feedback~\cite{distefano:meier1999}.

\section*{Results from Gran Sasso}
The experimental setup underground is in Hall~B of the Gran Sasso National Laboratories in Italy, where the  rock shielding, equivalent to 3800~meters of water, reduces the cosmic muon flux to about  1 $/ \mbox{m}^{2} / \mbox{h}$.  
The fast neutron flux has been measured as $\approx 10^{-6} / \mbox{cm}^{2} / \mbox{s}$.  The experiment is housed in a 3~storey building, erected around a Faraday cage to shield the apparatus from electromagnetic interferences.  Within the cage is a class~100 clean room protecting the cryostat from radioactive dust.  The cryostat itself has been designed and constructed to protect the detectors from vibrations as well as from radioactivity and electronic noises.  It reaches a base temperature of 6~mK.  Though its experimental volume is greater, so far it has housed up to four calorimeters.  They are mounted within 10~mm of each other 
and with no interposed material (other than small nibs at the corners of the crystals)
so as to make them self-shielding and to allow the study of coincidences between them.  More details of this setup are available in Ref.~\cite{distefano:CRESST1999}.

First results from the detectors were puzzling.  
The experimental rate was  orders of magnitude higher than expected, despite all the pulses having the  shape anticipated from calibrations with a \CoF\ source.  Moreover, the spectrum showed no lines or structures, the time distribution of events did not follow a Poisson distribution, 
and there were no coincidences between crystals.

Several attempts were made to understand this enigma. For instance, the detectors were isolated mechanically with a spring suspension, and the electrical circuits were modified.  These changes improved the stability of the setup to external perturbations, but did not affect the overall background significantly.  The next step was to install a special calorimeter consisting of a standard 262~g sapphire absorber with two tungsten films on it.  
\begin{figure}[t!] 
\centering
	\begin{tabular}{cc}
		\mbox{\epsfig{file=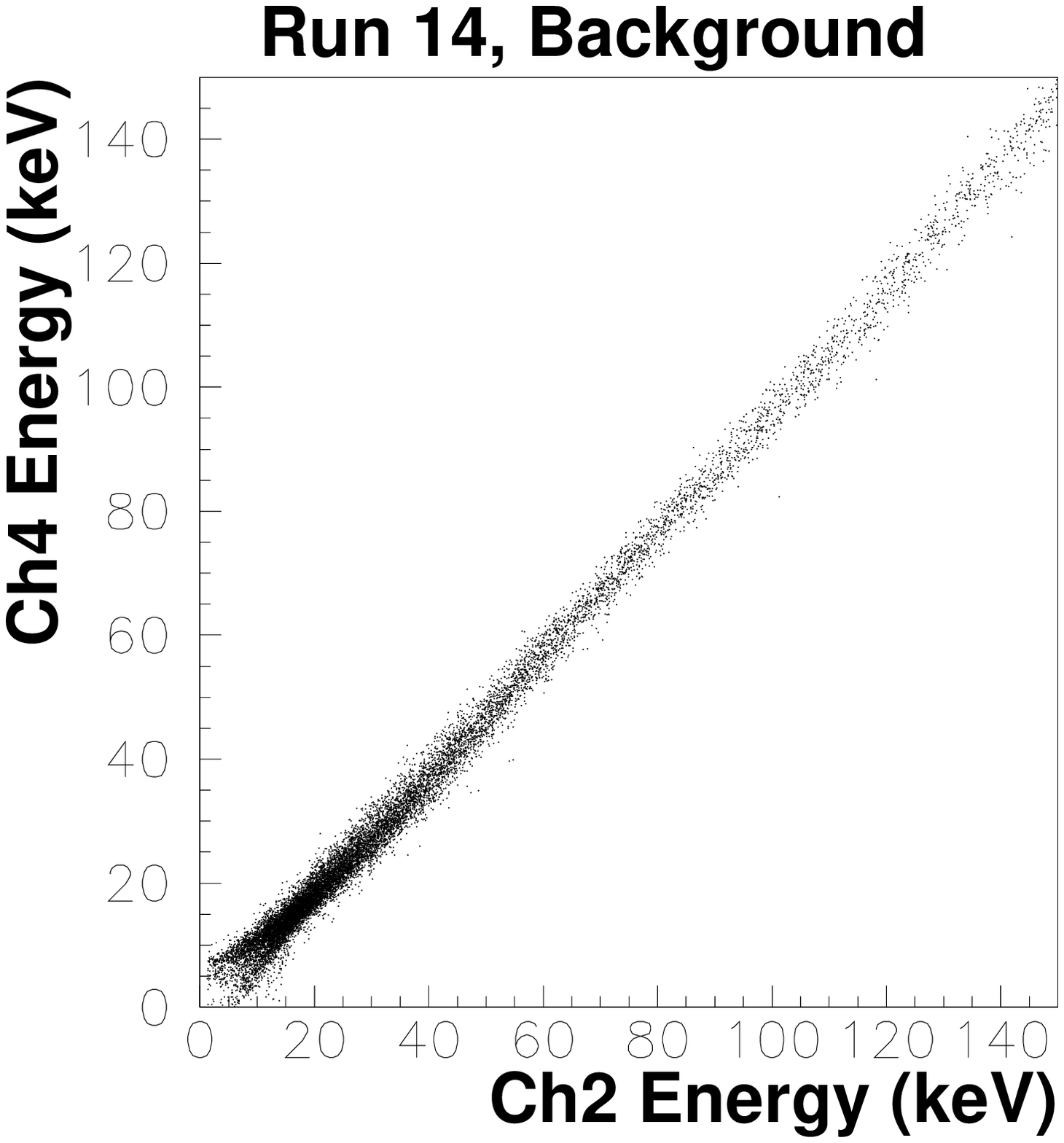,width=0.5\linewidth}} & 
		\mbox{\epsfig{file=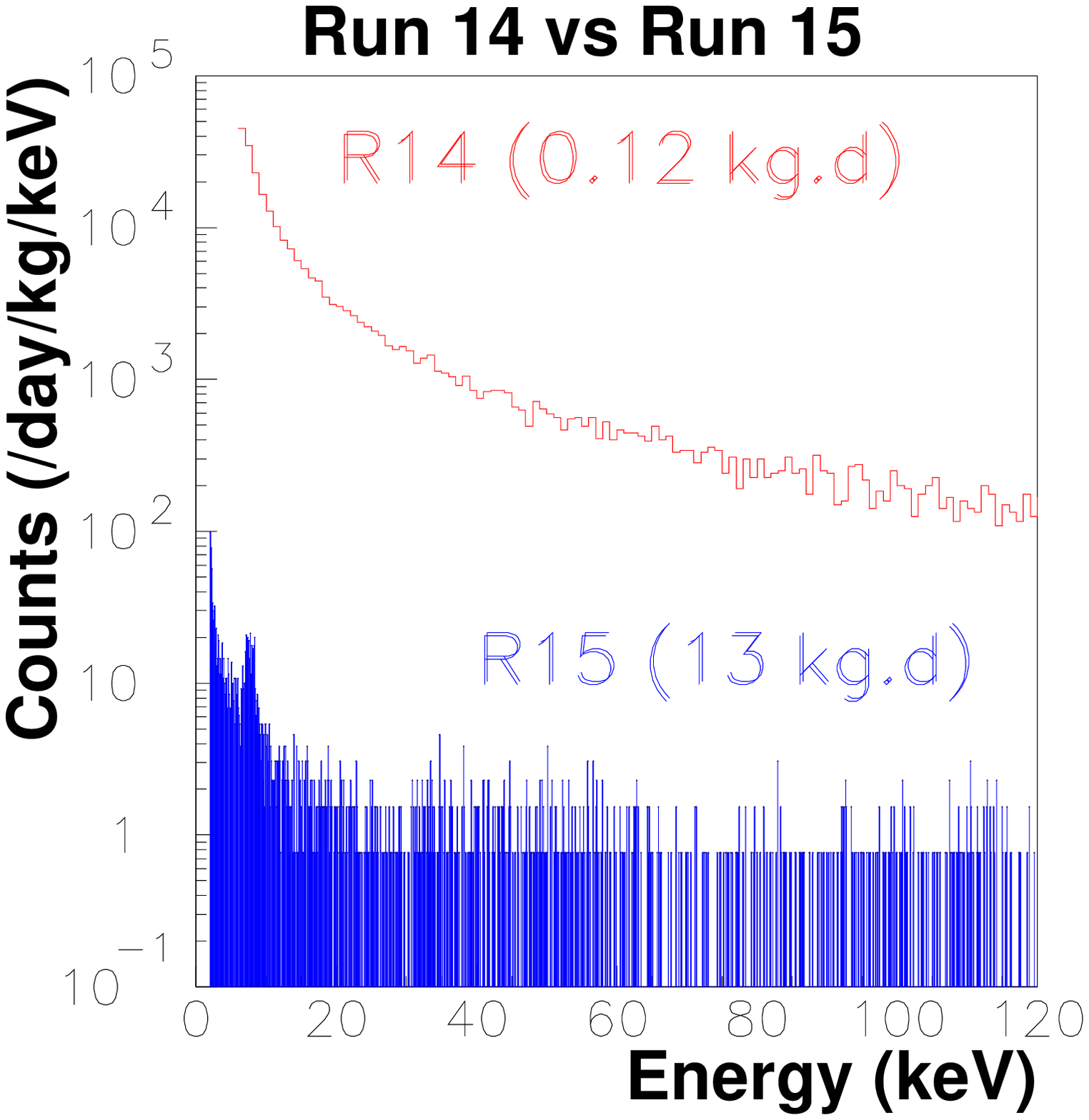,width=0.5\linewidth}}
	\end{tabular}
\caption{Left, pulses, as seen by both tungsten films on a common crystal, before removal of source of stress on crystal.  All pulses are found to be coincident and to be of same energy within the resolution of the detector, and thus originate within the absorber.  
The agreement between the two channels is also a check of the  consistency of the methods used.
Right, typical energy spectra before and after modifying crystal mounting.}
\label{distefano:fig:InternalCoincidences}
\end{figure}
All pulses were found to be coincident between these two sensors and of same energy  (Fig.~\ref{distefano:fig:InternalCoincidences}), demonstrating  that the background originated within the sapphire crystal itself rather than in the tungsten films or in the electrical circuitry.  
The cause was finally traced back to slight fracturing of the sapphire crystal under pressure from the sapphire balls maintaining it in place.  
Over the course of the run, a network of micro-cracks gradually develops, releasing mechanical energy as it spreads. 
These micro-cracks have since been observed under microscope.

To alleviate this problem, the crystals were next mounted using softer and blunter Delrin stubs rather than sapphire balls.  The improvement in the background is noticeable (Fig.~\ref{distefano:fig:InternalCoincidences}), and surpasses three orders of magnitude at low energies.  In a run lasting several months, trigger efficiency has been studied by means of the heater pulses, and is found to rise to 100~\% by 2~keV.  
This value is taken as a conservative threshold henceforth. 
 Above it, the time distribution of events is Poissonian.  
A structure is present in the spectrum at around 8~keV.  This could be due to copper and/or nickel X-rays, and is under investigation.  
Above this contamination, the rate is of the order of 1~count/d/kg/keV. 
As a whole, the spectrum is an improvement over previous smaller sapphire devices of other groups~\cite{distefano:EDELWEISS1996,distefano:ROSEBUD2000}. 
Limits in terms of WIMPs are in the process of being inferred.

\section*{Development of Scintillating Calorimeters}
Further progress requires being able to differentiate between the electron recoils caused by the photon and electron background, and the nuclear recoils expected to be induced by the neutron background and the WIMPs.  
Other groups have achieved this
using a simultaneous measure of charge and heat in semiconducting calorimeters~\cite{distefano:spooner1991,distefano:shutt1992,distefano:berge1999}.  
The CRESST group is developing an alternative technique, which involves measurement of scintillating light and phonons, 
using a  principal calorimeter made of a scintillating crystal which yields the phonon signal and emits photons read by a very sensitive secondary sapphire calorimeter (scintillation signal).  
Efficiency is enhanced by placing both calorimeters in a light collector.
The proof of principle experiment with a rectangular parallelepiped scintillator~\cite{distefano:meunier1999} 
has demonstrated a rejection of electron recoils greater than 99.7~\% for a 95~\% acceptance of nuclear recoils  above 15~keV,  
and no reduction in performance in the case of surface events.  
A scaled-up device is in preparation, the challenge being that in the proof of principle experiment, 
a mere 0.8~\% of the energy deposited as electron recoils in the main crystal was converted to scintillation and eventually read in the secondary calorimeter.  
For a given material and type of recoil, among the factors which may influence this fraction are the shape and size of the scintillator, 
the efficiency of the light collector, and the size and efficiency of the light detector.

A 300~g \CaW\ crystal of the readily available cylindrical geometry is being prepared as the main scintillator.  
It was initially feared that in such a shape, internal reflections would lead to a reduction of  escaping light as the radial position of the interaction increased.  
This was tested by exposing the bottom of the crystal (itself of 40~mm height and diameter) to a \CoS\ source supplying  1.17 and 1.33~MeV photons with interaction lengths in the crystal of between 2.5 and 3~cm.  The resulting light spectrum showed that the two peaks were well enough resolved 
($\Delta E_{FWHM} / E \approx$ 15~\%) 
for one to be able to conclude that the expected bulk effect is in fact not significant.  
This is interpreted to mean that scattering on impurities in the crystal and non specular refraction out of the crystal at surface irregularities dominate absorption within the crystal.
For the tungsten phonon sensor, transition temperatures of 40~mK have repeatedly been reached on small \CaW\ substrates.  
The evaporation process is being studied to make the 15~mK transitions obtained in some instances reproducible.
The higher transition temperature may already provide a better sensitivity than the glued film used for the demonstration experiment.  
An alternative solution, proximity effect Ir-Au films~\cite{distefano:nagel1994}, is under investigation.

The second thrust of development is the light collector.  
The present design  uses special reflective foil~\cite{distefano:foil2000} wrapped into a cylinder concentric to the scintillating crystal, 
with two diffusive Teflon end-caps, one of which holds the light detector.  
Using the same light detector as in the demonstration experiment, 
this has yielded a new ratio of light detected over energy deposited  in the scintillator of 0.68~\% for electron recoils.  
Therefore, though the scintillator mass has been increased fiftyfold to 300~g, this setup still gives 85~\% of the light of the smaller experiment.

The third component being optimized is the light detector, which will be crucial for the threshold of the scaled-up device.  
Its design presents an apparent dilemma in that sensitivity of the calorimeter requires its size to be small, whereas efficiency of light collection requires its surface to be large.  
A $40 \times 40 \times 0.5 \mbox{ mm}^3$ sapphire calorimeter (sixteen times the  effective surface area of the demonstration light detector)  with a tungsten film has provided encouraging results : resolution FWHM of 150~eV  at 6~keV and baseline noise FWHM of 25~eV (in all likelihood dominated by the readout, which is noisier than that at Gran Sasso).  
Moreover, response to a collimated source shining on three different points of the calorimeter was homogeneous.  
With a slightly smaller and more sensitive light detector, the scintillator and light collector described above, it is hoped to 
duplicate the proof of principle rejection factor above phonon energies of 15~keV.
Assuming those performances and a background equal to that obtained with the sapphire, and inducing mainly electron recoils, 
300~g of \CaW\  exposed for three months would have a sensitivity comparable to the best spin-independent limits~\cite{distefano:CDMS2000}, 
thanks to the background suppression and the large coherent cross section of tungsten.

\section*{Conclusion and Outlook}
The CRESST dark matter experiment has demonstrated the feasibility of running massive cryogenic detectors in stable conditions underground for several months.  
A main source of background, the micro-cracking of the crystals under stress, has been identified and eliminated.  
This leaves a background rate which is now competitive with that of other experiments, and is probably dominated by photons or electrons.

Scintillating calorimeters being developed by the collaboration will  reduce  the impact of part of this background. 
In addition to a two orders of magnitude suppression of the background afforded by the discrimination between electron and nuclear recoils, 
the spin-independent sensitivity will be much improved through the higher coherent cross section of tungsten.  
Indeed, assuming a phonon threshold of 15~keV and a background rejection of 99.7~\% (both reached on a smaller-scale experiment), 
and assuming the same exposure time and background as those of  the sapphire devices, 
a 300~g \CaW\ scintillating calorimeter would have a sensitivity on a par with current best 
limits~\cite{distefano:CDMS2000}.
This 300~g device will be a stepping-stone to  a 10~kg array of 33 such scintillating calorimeters, for which planning and development are already underway.  
With adequate neutron shielding, this next major phase of the experiment will cover more of supersymmetric parameter space~\cite{distefano:CRESSTProp2000}.

\section*{Acknowledgments}
This work has been supported by EU-TMR Network for Cryogenic Detectors ERB-FMRX-CT98-0167
 and SFB 375-95 f\"ur Astro-Teilchenphysik der DFG.

\bibliographystyle{abbrv}

\end{document}